\documentclass[showpacs]{revtex4}
\usepackage{amsmath}
\usepackage{amssymb}
\usepackage{graphicx}

\begin{document}

\title{ Relativistic polytropic spheres embedded in a chameleon scalar field}

\author{
Vladimir Folomeev,$^{1,2}$
\footnote{Email: vfolomeev@mail.ru}
Douglas Singleton$^{3}$
\footnote{Email: dougs@csufresno.edu}
}
\affiliation{
$^1$Institut f\"ur Physik, Universit\"at Oldenburg, Postfach 2503
D-26111 Oldenburg, Germany
\\ 
$^2$Institute of Physicotechnical Problems and Material Science of the NAS
of the
Kyrgyz Republic, 265 a, Chui Street, Bishkek, 720071,  Kyrgyz Republic \\
$^{3}$ Physics Department, California State University, Fresno, Fresno, California  93740-8031, USA\\
}

\begin{abstract}
In this paper we investigate gravitationally bound, spherically symmetric equilibrium configurations
consisting of ordinary (polytropic) matter nonminimally coupled to an external chameleon scalar field.
We show that this system has static, regular, asymptotically flat general relativistic solutions.
The properties of these spherical configurations, such as total mass, distribution of matter, and size, depend strongly
on the surrounding scalar field. The mass is found
in terms of the parameter $\sigma$, which is proportional to the central mass density of the ordinary matter.
We perform a stability analysis of these spherical solutions and find
an upper bound for $\sigma$ where dynamical instability occurs.
\end{abstract}

\maketitle

\section{Introduction}
Scalar fields have been applied to a wide range of model building in cosmology and astrophysics. At the cosmological scale
scalar fields have been used to model inflation and the late time acceleration due to dark energy. At the galactic scale
scalar fields have been tried as models for dark matter (for a review with extensive references to scalar field models
of dark energy and dark matter, see \cite{sahni:2004}). At the astrophysical scale there are
boson stars (see the reviews \cite{Jetzer:1991jr,Schunck:2003kk}). Boson stars are stellar size or smaller,
self-gravitating configurations of scalar fields. The scalar fields may be self-interacting or not, and one can
include a fluid which interacts with the scalar field either only gravitationally \cite{Henriques:1989ar,Henriques:1989ez}
or through  direct coupling \cite{Lee:1986tr}. The properties of these boson stars strongly depend
on the type of scalar fields of which they are composed as well as how these fields
interact with each other and the surrounding matter/fluid.

In scalar-tensor gravitational theories there is also the possibility of constructing
mixed configurations of a scalar field plus ordinary matter. In such theories,
a conformal transformation from the original Jordan frame to the Einstein frame is always accompanied
by the appearance of a nonminimal coupling between ordinary matter and a scalar field, which
leads to new features not present in general relativity.
In particular, this nonminimal coupling between the matter of a neutron star and the scalar field
can lead to the effect of spontaneous scalarization of the neutron star \cite{Damour:1993hw,Damour:1996ke}.
In turn, this leads to a redistribution of the scalar field inside the
neutron star, which   has a significant influence on the process of the neutron star's collapse  \cite{Novak:1997hw}.

In cosmology, this ability of scalar fields to interact nongravitationally with other fields has been
used within the framework  of chameleon cosmology  \cite{Khoury:2003aq,Khoury:2003rn,Brax:2004qh},
where a chameleon scalar field $\phi$ interacts directly with ordinary matter through a conformal coupling of the
form $e^{\alpha \phi}$.  The name ``chameleon scalar field'' was suggested in \cite{Khoury:2003aq}
since the properties of the scalar field, such as its  mass, depend sensitively on the environment in which
the scalar field finds itself. The idea that the properties of a scalar field could be influenced
by the environment/matter surrounding the scalar field was studied earlier,
in particular,
 in the papers \cite{Ellis:1989as,Mota:2003tc}
where the interaction between matter and the scalar field was used to model a dependence of
fundamental  coupling  constants on the local environment.
The possibility of having scalar fields interact directly with ordinary matter has been used in~\cite{Brax:2004qh}
to describe the present accelerated expansion of the Universe.
These cosmological models were developed further in \cite{Farajollahi:2010pk}, where the authors choose a generalized expression
for the nonminimal coupling between the scalar field and the matter having the form $f(\phi) L_m$,
with $f(\phi)$ being an arbitrary coupling function and $L_m$ being the Lagrangian of ordinary matter.
In this case, for certain parameters of the model,
the numerical calculations of \cite{Farajollahi:2010pk} showed that it was possible to describe
the evolving
Universe with a contraction phase, a subsequent bounce phase, and finally a phase of accelerated expansion.
One can also find analytical solutions \cite{Cannata:2010qd} within the framework of the cosmological
model of \cite{Farajollahi:2010pk}. These analytical solutions describe the evolution in time of the
chameleon scalar field.

It is also
possible to consider the effect of a chameleon scalar field at astrophysical scales rather than cosmological scales.
In \cite{Dzhunushaliev:2011ma} the  influence of a chameleon field on the structure of a star composed of a
polytropic fluid nonminimally coupled to a chameleon scalar field was studied. It was found that
for special choices of the scalar field potential energy and the coupling function $f(\phi)$, one could
obtain static, regular, asymptotically flat solutions in both relativistic and
nonrelativistic limits. However, a preliminary stability analysis of the solutions found in
\cite{Dzhunushaliev:2011ma} indicated that they were not stable in that the matter of the solutions
would either collapse or disperse. In \cite{Folomeev:2011uj} it was shown that taking an
isothermal fluid coupled to a chameleon scalar field led to both stable and unstable static solutions, and
for the  unstable solutions,  the problem of the gravitational collapse of the nonrelativistic configurations
has been considered.

In this paper we continue the study of the influence of a chameleon scalar field on the structure of
polytropic spheres begun in \cite{Dzhunushaliev:2011ma}.  We consider a relativistic,
polytropic, spherically symmetric fluid embedded in an external  chameleon scalar field, which is distributed
homogeneously and isotropically over the Universe.  When polytropic matter is placed in such a homogeneous background,
it begins to interact with the background scalar field both gravitationally and through the nonminimal coupling.
This results in a change of the inner structure of the polytropic star depending on the properties of the surrounding
chameleon scalar field. In general, the scalar field will evolve over cosmological time, and this will
lead to a time dependence for the term $f(\phi) L_m$, which in turn can have a strong influence on the structure of the
compact fluid configuration which is embedded in the scalar field. In this paper we restrict ourselves to
only a simplified version of the problem, when we just study changes in the structure of a {\it static} polytropic, fluid
sphere embedded in an external {\it dynamic} chameleon scalar field. For the time variation of the scalar field we
take the time dependence suggested in \cite{Cannata:2010qd} (see below in Sec.~\ref{num_res}).

The paper is organized as follows: In Sec.~\ref{gen_equations_cham_star} the general equations
describing an equilibrium configuration consisting of a real scalar field coupled to a perfect
polytropic fluid are derived. In Sec.~\ref{static_sol_exp} these
equations are written for a particular case when the coupling function $f(\phi)$ is chosen
to be the exponential one. In Sec.~\ref{num_res} we give the results of the numerical calculations
 for  this choice of nonminimal coupling function.
In Sec.~\ref{stab_gen_exp} we address the issue of the stability of the
obtained equilibrium configurations against small radial perturbations.
Finally, in Sec.~\ref{concl_exp} we summarize the main results.

\section{Derivation of equations for hydrostatic equilibrium configurations}
\label{gen_equations_cham_star}

As discussed in the Introduction we consider a gravitating system of a real scalar field coupled to a perfect fluid.
The Lagrangian for this system is
\begin{equation}
\label{lagran_cham_star}
L=-\frac{c^4}{16\pi G}R+\frac{1}{2}\partial_{\mu}\varphi\partial^{\mu}\varphi -V(\varphi)+f(\varphi) L_m ~.
\end{equation}
Here $\varphi$ is the real scalar field with the potential $V(\varphi)$; $L_m$ is the Lagrangian of the perfect
isotropic fluid i.e. a fluid with only one radial pressure; $f(\varphi)$ is some function describing the
nonminimal interaction between the fluid and the scalar field. The case $f=1$ corresponds to no direct coupling between
the fluid and scalar field, but even in this case the two sources are still coupled via gravity.

The Lagrangian for an isentropic  perfect fluid has the form $L_m=p$ \cite{Stanuk1964,Stanuk}.
Using this Lagrangian, the corresponding energy-momentum tensor is (details are given in the appendix of
\cite{Dzhunushaliev:2011ma})
 \begin{equation}
\label{emt_cham_star}
T_i^k=f\left[(\varepsilon+p)u_i u^k-\delta_i^k p\right]+\partial_{i}\varphi\partial^{k}\varphi
-\delta_i^k\left[\frac{1}{2}\partial_{\mu}\varphi\partial^{\mu}\varphi-V(\varphi)\right] ~,
\end{equation}
where $\varepsilon$ and $p$ are the total energy density and the pressure of the fluid, and $u^i$ is the four-velocity.
Bearing in mind that we will consider below small motions of the matter in the radial direction, we take the
time-dependent metric of the form
 \begin{equation}
\label{metric_sphera}
ds^2=e^{\nu} d(x^0)^2-e^{\lambda}dr^2-r^2d\Omega^2,
\end{equation}
where $\nu$ and $\lambda$ are functions of the radial coordinate $r$ and time coordinate  $x^0=c\, t$,
and $d\Omega^2$ is the metric on the unit two-sphere.
The $(_0^0)$, $(_1^1)$ and $(_0^1)$ components of the Einstein equations for the metric
\eqref{metric_sphera} and the energy-momentum tensor \eqref{emt_cham_star} are then given by
\begin{eqnarray}
\label{Einstein-00_cham_star}
&&G_0^0=-e^{-\lambda}\left(\frac{1}{r^2}-\frac{\lambda^\prime}{r}\right)+\frac{1}{r^2}
=\frac{8\pi G }{c^4}T_0^0,
 \\
\label{Einstein-11_cham_star}
&&G_1^1=-e^{-\lambda}\left(\frac{1}{r^2}+\frac{\nu^\prime}{r}\right)+\frac{1}{r^2}
=\frac{8\pi G }{c^4}T_1^1,\\
\label{Einstein-10_cham_star}
&&G_0^1=-e^{-\lambda}\frac{\dot{\lambda}}{r}
=\frac{8\pi G }{c^4}T_0^1.
\end{eqnarray}
In the above equations, the prime and dot denote differentiation with respect to $r$ and $x^0$,
respectively.

The equation for the scalar field $\varphi$ coming from the Lagrangian \eqref{lagran_cham_star} is
 \begin{equation}
\label{sf_eq_gen}
\frac{1}{\sqrt{-g}}\frac{\partial}{\partial x^i}\left[\sqrt{-g}g^{ik}\frac{\partial \varphi}{\partial x^k}\right]=
-\frac{d V}{d \varphi}+L_m \frac{d f}{d \varphi}.
\end{equation}

Not all of the Einstein field equations are independent because of the
law of conservation of energy and momentum,
$T^k_{i;k}=0$. Taking the $i=1$ component of this equation gives
\begin{equation}
\label{conserv_1_cham_star}
\frac{\partial T^0_1}{\partial x^0}+
\frac{\partial T^1_1}{\partial r}+
\frac{1}{2}(\dot\nu+\dot\lambda)T^0_1+
\frac{1}{2}\left(T_1^1-T_0^0\right)\nu^\prime+\frac{2}{r}\left[T_1^1-\frac{1}{2}\left(T^2_2+T^3_3\right)\right]=0.
\end{equation}

Thus we have five unknown functions: $\nu, \lambda, \varphi, \varepsilon$, and $p$.
Keeping in mind
that $\varepsilon$ and $p$ are related by some equation of state,  there
are only four unknown functions.
In this section we consider static equilibrium configurations so that Eq.~\eqref{Einstein-10_cham_star} is  trivially satisfied,
leaving four equations: the two Einstein equations \eqref{Einstein-00_cham_star} and \eqref{Einstein-11_cham_star},
the scalar field equation \eqref{sf_eq_gen}, and  the
hydrostatic equilibrium equation  \eqref{conserv_1_cham_star}.
Using the energy-momentum tensor \eqref{emt_cham_star}, the right-hand sides of
Eqs.~\eqref{Einstein-00_cham_star} and \eqref{Einstein-11_cham_star} take the form
\begin{eqnarray}
\label{T00_stat}
&&T_0^0=f \varepsilon+\frac{1}{2}e^{-\lambda}\varphi^{\prime 2}+V(\varphi),
 \\
\label{T11_stat}
&&T_1^1=-f p-\frac{1}{2}e^{-\lambda}\varphi^{\prime 2}+V(\varphi).
\end{eqnarray}
In turn, the scalar field equation \eqref{sf_eq_gen} [with the perfect fluid Lagrangian of the form $L_m=p$, and with the metric
\eqref{metric_sphera}] is
\begin{equation}
\label{sf_cham_star}
\varphi^{\prime\prime}+\left[\frac{2}{r}+\frac{1}{2}\left(\nu^\prime-\lambda^\prime\right)\right]\varphi^\prime=
e^{\lambda}\left(\frac{d V}{d \varphi}-p\frac{d f}{d\varphi}\right)~.
\end{equation}

Next, taking into account the corresponding components of the energy-momentum tensor from \eqref{emt_cham_star}:
$$
T_2^2=T_3^3=-f p+\frac{1}{2}e^{-\lambda}\varphi^{\prime 2}+V(\varphi) ~,
$$
and $T_0 ^0$, $T_1 ^1$ from \eqref{T00_stat}, \eqref{T11_stat} and using
\eqref{sf_cham_star}, allows us to write \eqref{conserv_1_cham_star}, in the static case, as
\begin{equation}
\label{conserv_2_cham_star}
\frac{d p}{d r}=-\frac{1}{2}(\varepsilon+p)\frac{d\nu}{d r}.
\end{equation}

Finally, we need an equation of state for the fluid. We use the following
parametric relation between the pressure and total  energy density of the fluid \cite{Tooper2}:
\begin{equation}
\label{eqs_cham_star}
p=K \rho_{g}^{1+1/n}, \quad \varepsilon = \rho_g c^2 +n p.
\end{equation}
The constant $n$ is called the polytropic index and is related to the specific heat ratio $\gamma$ via
$n=1/(\gamma-1)$, $\rho_g$ is the rest-mass density of the fluid
consisting of a gas of noninteracting particles,
and $K$ is  an arbitrary constant
whose value depends on the properties of the fluid under consideration.

Introducing the new variable $\theta$ defined as in \cite{Zeld},
\begin{equation}
\label{theta_def}
\rho_g=\rho_{g c} \theta^n~,
\end{equation}
where $\rho_{g c}$ is the central density of the fluid, we may rewrite the pressure from Eq.~\eqref{eqs_cham_star} in the form
\begin{equation}
\label{pressure_fluid_theta}
p=K\rho_{g c}^{1+1/n} \theta^{n+1}.
\end{equation}
Using \eqref{eqs_cham_star} and \eqref{pressure_fluid_theta} in \eqref{conserv_2_cham_star} leads to
\begin{equation}
\label{conserv_3_cham_star}
2\sigma(n+1)\frac{d\theta}{d r}=-\left[1+\sigma (n+1)\theta\right]\frac{d\nu}{dr},
\end{equation}
with $\sigma=K \rho_{g c}^{1/n}/c^2=p_{c}/(\rho_{g c} c^2)$ and $p_{c}$ is the pressure of the fluid at the center of the configuration.
This equation may be integrated to give $e^{\nu}$ in terms of $\theta$:
\begin{equation}
\label{nu_app}
	e^{\nu}=e^{\nu_c}\left[\frac{1+\sigma (n+1)}{1+\sigma (n+1)\theta}\right]^{2},
\end{equation}
where $e^{\nu_c}$ is the value of $e^{\nu}$ at the center of the configuration where $\theta=1$.
The integration constant $\nu_c$, corresponds to the value of $\nu$ at the center of the
configuration. It is fixed by requiring $e^{\nu}=1$ at infinity i.e. that the
space-time be asymptotically flat.

Thus, the gravitating system of a real scalar field interacting with a perfect fluid
is characterized by three unknown functions -- $\lambda, \theta$, and $\varphi$. These three
functions are determined by the three equations \eqref{Einstein-00_cham_star}, \eqref{Einstein-11_cham_star}
and \eqref{sf_cham_star}, and also by the relation \eqref{nu_app}. We rewrite these equations
by introducing a new function $M(r)$,

\begin{equation}
\label{u_app}
e^{-\lambda}=1-\frac{2 G M(r) }{c^2 r}~.
\end{equation}
$M(r)$ is the mass of the configuration in the range $0 \leq r \leq R$,
where $R$ is the boundary of the fluid where $\theta=0$.
Using this function, Eqs.~\eqref{Einstein-00_cham_star} and \eqref{T00_stat} yield
\begin{equation}
\label{00_via_u}
\frac{d M}{d r}=\frac{4\pi}{c^2} r^2\left[f\varepsilon+\frac{1}{2}\left(1-\frac{2 G M(r) }{c^2 r}\right)\varphi^{\prime 2}+V\right].
\end{equation}
To avoid a singularity at the origin, one has to require that $M(0)=0$.

Since we will analyze the system of equations \eqref{Einstein-00_cham_star}, \eqref{Einstein-11_cham_star}
and \eqref{sf_cham_star} numerically, we introduce dimensionless variables
\begin{equation}
\label{dimless_xi_v}
\xi=A r, \quad v(\xi)=\frac{A^3 M(r)}{4\pi \rho_{g c}},
\quad \phi(\xi)=\left[\frac{4\pi G}{\sigma(n+1)c^4}\right]^{1/2}\varphi(r),
\quad \text{where} \quad A=\left[\frac{4\pi G \rho_{g c}}{\sigma (n+1) c^2}\right]^{1/2},
\end{equation}
where $A$ has the dimensions of inverse length. With this one can rewrite
\eqref{Einstein-00_cham_star} and \eqref{Einstein-11_cham_star} in the form
\begin{eqnarray}
\label{eq_v_app}
\frac{d v}{d\xi} &=& \xi^2\left\{f(1+n\sigma \theta)\theta^n+\frac{1}{2}\left[
1-2\sigma(n+1)\frac{v}{\xi}
\right]
\left(\frac{d\phi}{d\xi}\right)^2+\tilde{V}\right\}
,\\
\label{eq_theta_app}
\xi^2\frac{1-\frac{2\sigma(n+1)v}{\xi}}{1+\sigma(n+1)\theta}\frac{d\theta}{d\xi} &=&
\xi^3\left[-f\sigma\theta^{n+1}-
\frac{1}{2}\left[
1-2\sigma(n+1)\frac{v}{\xi}
\right]
\left(\frac{d\phi}{d\xi}\right)^2+\tilde{V}\right]-v ~,
\end{eqnarray}
where $\tilde{V}=V/(\rho_{g c} c^2)$ is the dimensionless potential energy of the field.

Next, using  \eqref{conserv_3_cham_star} and \eqref{u_app}, we rewrite \eqref{sf_cham_star} as follows:
\begin{eqnarray}
\label{eq_phi_dim_cham_star}
\frac{d^2 \phi}{d\xi^2}&+&\left\{\frac{2}{\xi}-\frac{\sigma(n+1)}{1+\sigma(n+1) \theta}
\left[\frac{d\theta}{d\xi}+\frac{1+\sigma (n+1)\theta}{1-\frac{2\sigma(n+1)v}{\xi}}\frac{1}{\xi}
\left(\frac{d v}{d\xi}-\frac{v}{\xi}\right)\right]\right\}\frac{d\phi}{d\xi} \nonumber\\
&&=\left[1-2\sigma(n+1)\frac{v}{\xi}\right]^{-1}\left(\frac{d \tilde{V}}{d\phi}-\sigma \theta^{n+1}\frac{d f}{d\phi}\right).
\end{eqnarray}
Thus, the static configuration under consideration is described by the three equations \eqref{eq_v_app}-\eqref{eq_phi_dim_cham_star}.

\section{Equilibrium configurations with $f=a e^{-b \phi}$}
\label{static_sol_exp}

Here we use the equations obtained in the previous section with the specific
choice of the coupling function $f$,
\begin{equation}
\label{fun_f_exp}
f=a e^{-b\phi},
\end{equation}
where $a, b$ are arbitrary constants. Such a form for the coupling function was used in
\cite{Farajollahi:2010pk} to study the evolution of the Universe within the framework of
chameleon cosmology. In this paper it was shown that choosing different values of the parameters $a, b$
gave different forms of the effective equation of state of matter filling the Universe.

Here we restrict ourselves to the simplest variant when the potential energy is absent i.e. $V(\phi)=0$.
In the absence of the nonminimal coupling, this corresponds to the case of a massless scalar field.
However, in the case being considered here one can see from the right-hand side of \eqref{eq_phi_dim_cham_star}
that the presence of the term with a nonminimal coupling leads to
an effective mass term of the form $\sigma \theta^{n+1} d f/d\phi$.
Our purpose is to study the influence that such a chameleon scalar field can have on the inner structure
of polytropic spherical configurations. In this case the total energy density of the system
$T_0^0$ from \eqref{emt_cham_star} takes the form (in units of $\rho_{g c} c^2$)
\begin{equation}
\label{tot_dens_poly_scalar_exp}
T_0^0=a e^{-b \phi}(1+n\sigma \theta)\theta^n+\frac{1}{2}\left[
1-2\sigma(n+1)\frac{v}{\xi}
\right]
\left(\frac{d\phi}{d\xi}\right)^2.
\end{equation}
Because of the spherical symmetry of the problem, one needs to require that
$d\phi/d\xi=0$ at the center. To ensure the positiveness of the energy density over the whole volume of the sphere,
it is necessary to choose $a>0$. Furthermore, from \eqref{eq_v_app}-\eqref{eq_phi_dim_cham_star}
the parameter $a$ can be absorbed by introducing the rescaling $\sqrt{a} \xi = \bar{\xi}$ and $\sqrt{a} v = \bar{v}$. Thus
we set $a=1$ in further calculations and drop the bar for simplicity. But we will bear in mind that
$a$ can always be restored by using the above rescaling, leading correspondingly to a change of the size and the mass
of the configuration under consideration.
Taking all this into account, we can rewrite \eqref{eq_v_app}-\eqref{eq_phi_dim_cham_star} as follows:
\begin{eqnarray}
\label{eq_v_exp}
\frac{d v}{d\xi} &=& \xi^2\left\{ e^{-b \phi}(1+n\sigma \theta)\theta^n+\frac{1}{2}\left[
1-2\sigma(n+1)\frac{v}{\xi}
\right]
\left(\frac{d\phi}{d\xi}\right)^2\right\}
,\\
\label{eq_theta_exp}
\xi^2\frac{1-\frac{2\sigma(n+1)v}{\xi}}{1+\sigma(n+1)\theta}\frac{d\theta}{d\xi} &=&
\xi^3\left[- e^{-b \phi}\sigma\theta^{n+1}-
\frac{1}{2}\left[
1-2\sigma(n+1)\frac{v}{\xi}
\right]
\left(\frac{d\phi}{d\xi}\right)^2\right]-v ~,\\
\label{eq_phi_dim_cham_star_exp}
\frac{d^2 \phi}{d\xi^2}&+&\left\{\frac{2}{\xi}-\frac{\sigma(n+1)}{1+\sigma(n+1) \theta}
\left[\frac{d\theta}{d\xi}+\frac{1+\sigma (n+1)\theta}{1-\frac{2\sigma(n+1)v}{\xi}}\frac{1}{\xi}
\left(\frac{d v}{d\xi}-\frac{v}{\xi}\right)\right]\right\}\frac{d\phi}{d\xi} \nonumber\\
&&=\left[1-2\sigma(n+1)\frac{v}{\xi}\right]^{-1} b\, \sigma  e^{-b \phi}\theta^{n+1}.
\end{eqnarray}
These equations are to be solved for given $\sigma$, $n$, and $b$ subject to the boundary conditions in the vicinity
of the center of the configuration $\xi=0$,
\begin{equation}
\label{bound_all}
\theta \simeq \theta_0+\frac{\theta_2}{2}\xi^2, \quad v \simeq v_3 \xi^3, \quad
\phi \simeq \phi_0+\frac{\phi_2}{2}\xi^2,
\end{equation}
where  $\phi_0$ corresponds to the initial value of the scalar field
$\phi$, $\theta_0$ is normalized to be unity at the center, $\theta_0\equiv\theta(0)=1$,
the parameters $\theta_2, v_3$  are arbitrary, and the value of the coefficient
$\phi_2$ is defined from Eq.~\eqref{eq_phi_dim_cham_star_exp} as
$$
\phi_2=\frac{1}{3}\, b\, \sigma e^{-b \phi_0} \theta_0^{n+1}.
$$

It is convenient to rewrite the mass and radius of the configuration in terms of the constants $K, n$, and $\sigma$ (see \cite{Tooper2}).
By eliminating $\rho_{g c}$ from the expressions for $\xi$ and $v$ in \eqref{dimless_xi_v}, we obtain
\begin{eqnarray}
\label{r_struct}
&&r=R^*\sigma^{(1-n)/2}\xi, \quad R=R^*\sigma^{(1-n)/2}\xi_1,\\
\label{M_struct}
&&M(r)=M^*\sigma^{(3-n)/2}v(\xi), \quad M=M^*\sigma^{(3-n)/2}v(\xi_1),
\end{eqnarray}
where
\begin{eqnarray}
\label{r_struct_2}
&&R^*=(4\pi)^{-1/2}(n+1)^{1/2}G^{-1/2}K^{n/2}c^{1-n},\\
\label{M_struct_2}
&&M^*=(4\pi)^{-1/2}(n+1)^{3/2}G^{-3/2}K^{n/2}c^{3-n}.
\end{eqnarray}
The quantities $R^*$ and $M^*$ define the scales of the radius and mass.
Next, the metric component $g_{11}=-e^{\lambda}$ from \eqref{u_app} can be rewritten via the dimensionless
variables \eqref{dimless_xi_v} as follows:
\begin{equation}
\label{u_dimen}
e^{-\lambda}=1-2\sigma (n+1)\frac{v}{\xi}.
\end{equation}
From Eqs.~\eqref{nu_app} and \eqref{u_dimen} one sees that $e^{\nu}<1$ and $e^{\lambda} \geq 1$ along the radius
of the configuration, and that $e^{\lambda}=1$ at the center, while $e^{\nu}$ takes its minimum value $e^{\nu_c}$.
Outside the star, the solution approaches the Schwarzschild solution
\begin{equation}
\label{lambda_dim_ext}
e^{\nu}=e^{-\lambda}=1-2\sigma(n+1)\frac{v(\xi_1)}{\xi}
\end{equation}
with the total mass $M(R)\sim v(\xi_1)$ [see \eqref{dimless_xi_v}].

One may question the assertion that
the asymptotic behavior of \eqref{lambda_dim_ext} is Schwarzschild since in Sec.~\ref{num_res} we show that
the scalar field approaches a constant value in the region outside the star. Thus, it would seem that the solution should
be matched to de Sitter space-time in the external region. However, even though the scalar field
approaches a constant value in the region outside the star, the energy density of the scalar field goes to zero (see
the numerical results in Fig.~\ref{energ_fig}). This comes about since the scalar potential is zero, $V(\phi) =0$, so that
the only contribution to the scalar field energy density comes from the kinetic energy term which involves derivatives of the
scalar field. For a constant scalar field this kinetic energy term vanishes.

It will be shown below, in Sec.~\ref{num_res}, that the main part of the mass of the configuration under consideration
is concentrated inside the radius $\xi_1$. This allows one,
by equating  expression \eqref{lambda_dim_ext} to expression \eqref{nu_app} at the boundary of the configuration
$\xi=\xi_1$, where $\theta=0$, to find the approximate value of the metric function $\nu$ at the center of configuration via
\begin{equation}
\label{nu_cen}
e^{\nu_c}\approx \frac{1-2\sigma (n+1)v(\xi_1)/\xi_1}{\left[1+\sigma(n+1)\right]^2}.
\end{equation}

\subsection{Energies of the system}

In this subsection we review definitions \cite{Tooper2} for the proper energy, the total energy, and the binding
energy and apply these to the spherical configurations discussed in the previous sections. The purpose is to
get some insight into the stability of the configurations we will find numerically in
the next subsection.
 In particular,
if the binding energy of the configuration, as defined below, is negative, then the configuration is naively unstable against
dispersal of the fluid. A more rigorous, dynamical stability test will be studied in Sec.~\ref{stab_gen_exp}.
First, using \eqref{M_struct}, the total energy $E$ of the system, including the internal and gravitational energies, is given by
\begin{equation}
\label{total_energ}
E=M c^2=M^* c^2 \sigma^{(3-n)/2}v(\xi_1).
\end{equation}

Second, the proper energy density $E_0$ is the sum of the rest energy density $E_{0g}$ of the system,
\begin{equation}
\label{proper_energ_gas}
E_{0g}=M_{0g}c^2=4\pi \int_0^R
\left(f \rho_g c^2 +\frac{1}{2}e^{-\lambda}\varphi^{\prime 2}\right)
e^{\lambda/2} r^2 dr
\end{equation}
and its total microscopic kinetic energy,
\begin{equation}
\label{kinetic_energ_gas}
E_{0k}=4\pi \int_0^R f n p \,e^{\lambda/2} r^2 dr.
\end{equation}
Using expressions  \eqref{theta_def}, \eqref{pressure_fluid_theta}, and \eqref{u_dimen} for $\rho_g$, $p$, and $e^{\lambda}$,
respectively, we get
\begin{eqnarray}
\label{Eog_param}
&&E_{0g}=M^* c^2 \sigma^{(3-n)/2} u_g(\xi_1),\\
\label{Eok_param}
&&E_{0k}=M^* c^2 \sigma^{(3-n)/2} u_k(\xi_1),
\end{eqnarray}
where $u_g$ and $u_k$ are solutions of the equations
\begin{eqnarray}
\label{ug_eq}
&&\frac{d u_g}{d\xi}=\xi^2\frac{e^{-b \phi}\theta^n+1/2 \left[1-2\sigma (n+1)v/\xi\right](d\phi/d\xi)^2}
{\left[1-2\sigma (n+1)v/\xi\right]^{1/2}}, \quad u_g(0)=0,\\
\label{uk_eq}
&&\frac{d u_k}{d\xi}=\frac{n\sigma e^{-b \phi}\xi^2\theta^{n+1}}
{\left[1-2\sigma (n+1)v/\xi\right]^{1/2}}, \hspace{3.6cm} u_k(0)=0,
\end{eqnarray}
and $u_g(\xi_1), u_k(\xi_1)$ are the values of these functions at the boundary $\xi=\xi_1$.

Finally, the binding energy (B.E.) is the difference between the energy of the unbound particles dispersed to
infinity and the total energy of the bound system,
\begin{equation}
\label{bind_enrg}
\text{B.E.}=E_{0g}-E.
\end{equation}
In terms of $u_g$ and $v$ the binding energy can be expressed as
\begin{equation}
\label{bind_enrg_param}
\text{B.E.}=M^* c^2 \sigma^{(3-n)/2}\left[ u_g(\xi_1)-v(\xi_1)\right].
\end{equation}
A negative binding energy indicates that the configuration is unstable against the dispersal of the star's
matter
to infinity. Below, we will use this condition to address the question of the stability of the configurations under consideration.

\subsection{Numerical results}
\label{num_res}

We solved the system of equations \eqref{eq_v_exp}-\eqref{eq_phi_dim_cham_star_exp} numerically using the boundary conditions
\eqref{bound_all}. We started the solutions  near the origin (i.e. near $\xi \approx 0$) and solved out to a point
$\xi = \xi_1$,  where the function $\theta$ became zero. Since our spherical fluid configuration
was taken to be embedded in an external, homogeneously distributed scalar field, we also needed to require that at the
boundary, $\xi = \xi_1$, the value of the varying scalar field from the inside of the star,
$\phi (\xi )$,
matched
the external value $\phi_{\text{ext}}$; i.e. we required $\phi(\xi _1 ) \simeq \phi_{\text{ext}}$.
Under these conditions the numerical calculations indicate that on the boundary of the configuration the derivative
$(d\phi/d\xi)_{\xi_1}$,  although small, is not strictly equal to zero.
It requires matching of the internal  solutions with the external ones having a nonzero scalar field energy density
that vanishes asymptotically. That is why  beyond the point $\xi = \xi_1$ we continued the numerical solutions with
only the gravitational and scalar fields while the fluid was set to zero.

For this purpose, we write down the Einstein equations of
\eqref{Einstein-00_cham_star} and \eqref{Einstein-11_cham_star},
and the scalar field equation of \eqref{sf_eq_gen}, without the fluid source, i.e.~$\theta=0$.
This leads to the following system of equations:
\begin{eqnarray}
\label{Einstein-00_ext}
&&-e^{-\lambda}\left(\frac{1}{r^2}-\frac{\lambda^\prime}{r}\right)+\frac{1}{r^2}
=\frac{4\pi G}{c^4} e^{-\lambda}\varphi^{\prime 2},
 \\
\label{Einstein-11_ext}
&&-e^{-\lambda}\left(\frac{1}{r^2}+\frac{\nu^\prime}{r}\right)+\frac{1}{r^2}
=-\frac{4\pi G}{c^4}e^{-\lambda}\varphi^{\prime 2},\\
\label{first_int_ext}
&&\varphi^{\prime 2}=\frac{D^2}{r^4}e^{\lambda-\nu},
\end{eqnarray}
where $D$ is an integration constant.
This system, by analogy with the transformations made above,
can be rewritten in terms of the dimensionless variables
$v(\xi), \phi(\xi)$, and $\nu(\xi)$ as follows:
\begin{eqnarray}
\label{eq_v_ext}
	\frac{d v}{d\xi} &=& \frac{1}{2}\frac{\bar{D}^2}{\xi^2}e^{-\nu},
 \\
\label{eq_nu_ext}
	\frac{d\nu}{d\xi} &=&
	\frac{1}{\xi}\left[
		\frac{1+\sigma (n+1)\frac{\bar{D}^2}{\xi^2}e^{-\nu}}{1-2\sigma (n+1) v/\xi}
	-1\right],
\\
\label{eq_phi_ext}
	\left(\frac{d\phi}{d\xi}\right)^2 &=&
	\frac{\bar{D}^2}{\xi^4}
	\frac{e^{-\nu}}{1-2\sigma (n+1) v/\xi}\,,
\end{eqnarray}
where the new dimensionless constant
$$
\bar{D}=\frac{4\pi G D}{\sigma(n+1) c^3} \sqrt{\rho_{g c}}
$$
is introduced. This constant can be defined from \eqref{eq_phi_ext} as
$$
\bar{D}^2=\xi_1^4 e^{\nu(\xi_1)}\left[1-2\sigma (n+1) v(\xi_1)/\xi_1\right]  [\phi^{\prime 2}]_{\xi_1}.
$$
The system  \eqref{eq_v_ext}-\eqref{eq_phi_ext} contains the parameter $\sigma$ as a trace
of the influence of the fluid on the external  solution.
The solution is to be sought beginning from the surface of the fluid
at $\xi=\xi_1$ by using, as the boundary conditions, the values of $v(\xi_1), \phi(\xi_1)$
obtained from the solution of  equations \eqref{eq_v_exp}-\eqref{eq_phi_dim_cham_star_exp}
and $\nu(\xi_1)$ from expression \eqref{nu_app} for the internal part of the configuration.
This allows one to determine the value of the integration constant $\nu_c$ from \eqref{nu_app}
by requiring $e^{\nu}$ to be equal to unity at infinity, providing asymptotic flatness of the space-time.
Thus, the complete solution for the configuration under consideration is derived
by matching of the internal fluid solutions given by Eqs.~\eqref{eq_v_exp}-\eqref{eq_phi_dim_cham_star_exp}
with the fluid-free, external solutions obtained from the system \eqref{eq_v_ext}-\eqref{eq_phi_ext}.

In principle, the total mass of the configuration under consideration is determined as a sum of the internal mass, defined by
expression \eqref{M_struct}, and of the mass associated with the external scalar field.
However, in the case of a massless scalar field, a more elegant way
is to define the total mass via the Komar mass. The latter, in general, is defined as~\cite{Wald}
$$
M_{K}=\frac{2}{c^2}\int_{\Sigma} \left(T_{a b}-\frac{1}{2}g_{a b}T\right)n^a \xi^b dV,
$$
where $n^a$ is a normal to $\Sigma$ and
$\xi^b$ is a timelike Killing vector.
For the static spherically symmetric configuration being considered here,
we find
$$
M_{K}=\frac{4\pi}{c^2}\int_0^R f(\varepsilon+3 p)e^{(\nu+\lambda)/2}r^2 dr.
$$
Note that here the integration is performed only in the range from  0 to $R$, where there is a nonzero contribution
associated with the fluid. Using the dimensionless variables
\eqref{dimless_xi_v}, and also taking into account
 \eqref{eqs_cham_star}-\eqref{pressure_fluid_theta}, this expression can be rewritten in a dimensionless form as
$$
v_{K}\equiv \frac{A^3}{4\pi \rho_{g c}} M_{K}=
\int_0^{\xi_1} e^{-b \phi}\left[1+\sigma(n+3)\theta\right]\theta^n e^{(\nu+\lambda)/2} \xi^2 d\xi.
$$

Comparing this {\it total mass} with the mass of the internal part of the configuration, defined by
\eqref{M_struct}, one can see the contribution to the total mass from the external scalar field.
The numerical calculations presented below indicate that the value of this
contribution can be up to 1\% in some cases, but generally is much smaller. From this it is clear that
the main part of the mass  is concentrated inside the radius
$\xi_1$ (or $R$). This, in turn, allows one to introduce some  effective radius of the configuration
which is defined as a radius where a certain percentage of mass is concentrated (see e.g. Sec.~II~D in \cite{Schunck:2003kk}).
In our case it is natural to choose the effective radius as a radius of the fluid, where more than 99\%
of the total mass of the configuration is concentrated.

\begin{figure}[t]
\begin{minipage}[t]{.49\linewidth}
  \begin{center}
  \includegraphics[width=7cm]{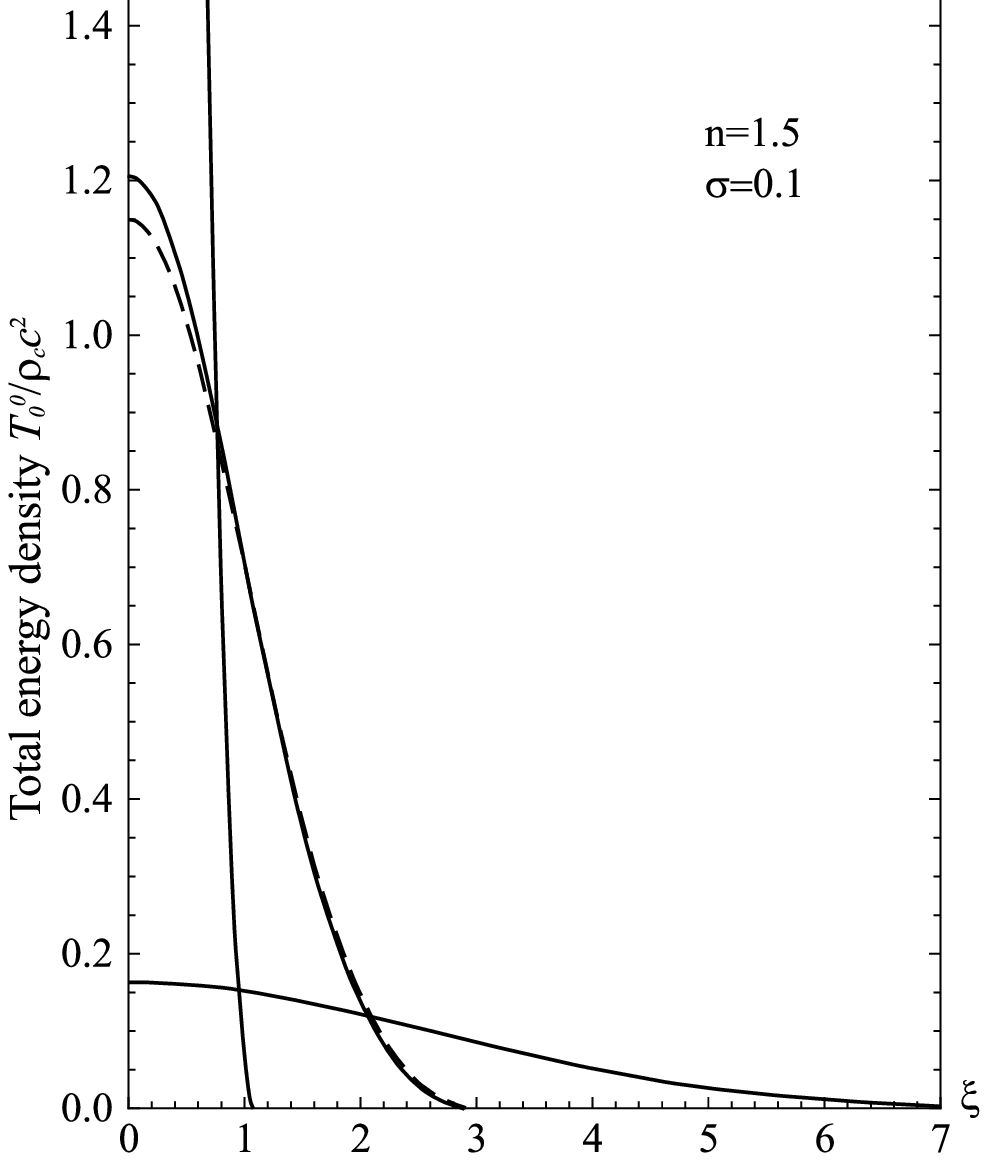}
  \end{center}
\end{minipage}\hfill
\begin{minipage}[t]{.49\linewidth}
  \begin{center}
  \includegraphics[width=7cm]{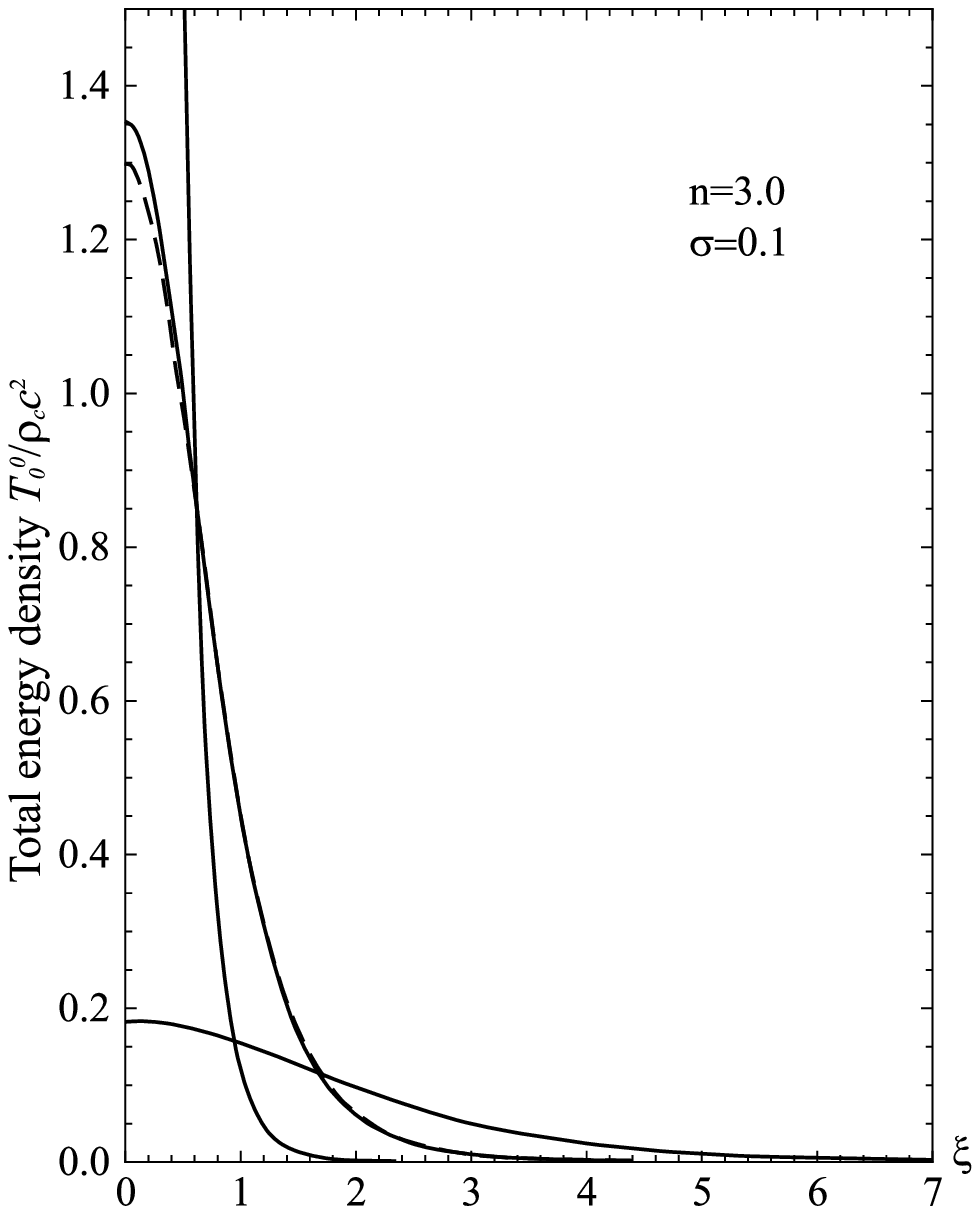}
  \end{center}
\end{minipage}\hfill
  \caption{\small The distributions of the total energy density in units of $\rho_c c^2$ from \eqref{tot_dens_poly_scalar_exp}
  for the different values of $\phi_{\text{ext}}=-2, 0, 2$, from top to bottom.
  At $\phi_{\text{ext}}=-2$ the central values of $T_0^0/\rho_c^2$ are as follows:
 for $n=1.5$, $T_0^0/\rho_c^2 = 8.91$;   for $n=3.0$, $T_0^0/\rho_c^2 = 10.01$  (not shown in the figure). For all curves $b=1$.
  The dashed line corresponds to the case without the chameleon scalar field considered in \cite{Tooper2}.
In the $n=1.5, \sigma=0.1$ case of the left figure the energy density smoothly (but rapidly) goes to zero beyond the surface of the star as occurs in the right 
figure. This is not seen explicitly due to the scale used in the left figure.
  }
 \label{energ_fig}
\end{figure}

The procedure for finding solutions is as follows: for given values of $b$, $\sigma$, and $n$, we choose
the central value of the scalar field, $\phi_0$, such that $\theta$ goes to zero and
$\phi_1 \simeq \phi_{\text{ext}}$ at some finite value of $\xi = \xi_1$.  As mentioned
above, $\xi = \xi _1$ corresponds to the surface of the star. If one assumes that the value of the scalar field
 $\phi_{\text{ext}}$ changes during the course of the evolution of the Universe,
then, in order to study a configuration embedded in such an evolving field,
it is necessary to specify how the function $\phi_{\text{ext}}(t)$ changes with time.
Obviously, this time dependence will be model dependent. Here we will use the dependence suggested in \cite{Cannata:2010qd}.
From some general considerations about cosmological evolution, the authors of \cite{Cannata:2010qd} showed  that the
chameleon scalar field may evolve according to the law
\begin{equation}
\label{cham_field_evol}
\phi(t) \sim \ln \frac{t}{t_R-t},
\end{equation}
where $t_R$ is the time when the Universe ends its evolution in the big rip singularity,
and the arbitrary factor in front of the logarithm we take to be of order unity.
In this model the value of the scalar field changes from $-\infty$ at the moment of the big bang up to $+\infty$ at $t=t_R$.
It crosses the zero value at $t = t_R/2$. This is called crossing the phantom divide since
the effective equation crossed from
 greater than $w_{eff}=-1$ to  less than $w_{eff}=-1$. A
field
with $w_{eff} < -1$ corresponds to phantom dark energy.  This moment corresponds approximately
to the present stage of evolution of the Universe. In addition, for the model of
\cite{Cannata:2010qd}, it is possible to choose parameters so that the coupling function $f(\phi)$ is positive definite.

\begin{figure}[t]
\centering
  \includegraphics[height=10cm]{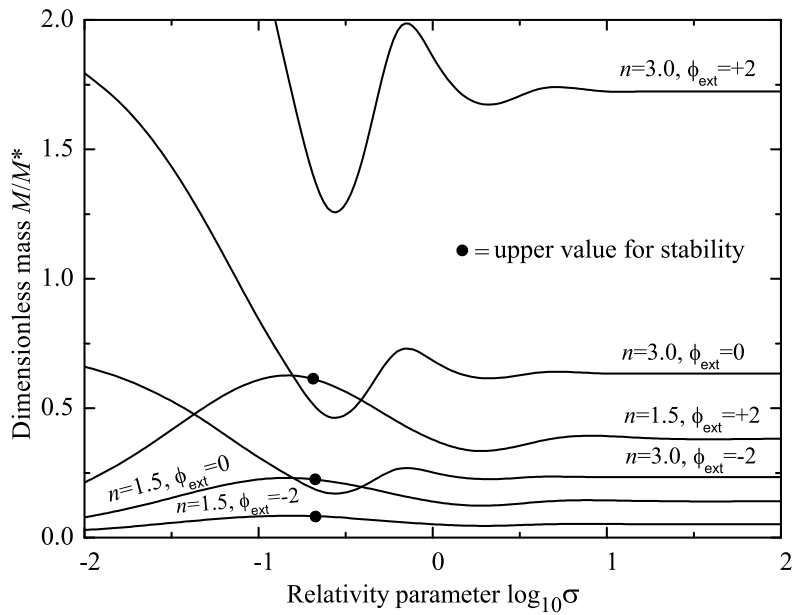}
\vspace{-1.cm}
\caption{The dimensionless mass given by Eq.~\eqref{M_struct} versus $\log_{10} \sigma$ for $n=1.5, 3.0$; $b=1$;
 and $\phi_{\text{ext}}=-2, 0, 2$. The dots represent the upper value for stability at
 $\log_{10}\sigma_{cr}\simeq -0.69$ for all $\phi_{\text{ext}}$.}
\label{fig_mass_exp}
\end{figure}

\begin{figure}[t]
\centering
  \includegraphics[height=10cm]{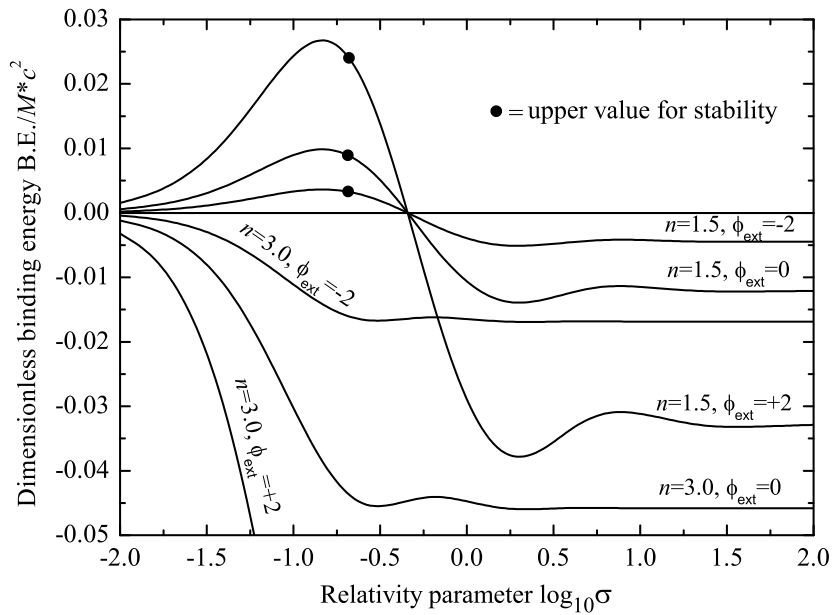}
\vspace{-1.cm}
\caption{The dimensionless binding energy given by Eq.~\eqref{bind_enrg_param} versus
$\log_{10} \sigma$ for $n=1.5, 3.0$; $b=1$; and $\phi_{\text{ext}}=-2, 0, 2$. All the binding energies
eventually become negative, but for $n=1.5$ there is first a peak in binding
energy after which instability occurs. For $n=3.0$ the binding energy is always negative.
The dots represent the upper value for stability at $\log_{10}\sigma_{cr}\simeq -0.69$ for all $\phi_{\text{ext}}$.}
\label{fig_bin_energy_exp}
\end{figure}

For the same values of the parameters  $b$, $\sigma$, and $n$,
the central value of the scalar field, $\phi_0$, will change depending on the value of $\phi_{\text{ext}}$.
Obviously, this will effect both the radial matter distribution of the star and the star's total mass.
In Figs.~\ref{energ_fig}-\ref{fig_bin_energy_exp} the results of the numerical calculations for the three values of
$\phi_{\text{ext}}=-2, 0, 2$ are presented (hereafter we choose $b=1$).
From \eqref{cham_field_evol} the field values $\phi_{\text{ext}}=-2, 0, 2$ correspond to the moments of
time $t_{-2}\approx 0.12 t_R$, $t_{0}=0.5 t_R$, and $t_{2}\approx 0.88 t_R$.
The time $t_0$ corresponds, approximately, to the current stage of the evolution of the Universe,
 $t_{-2}$ to the early Universe, and $t_{2}$ to the future Universe near the big rip singularity. Thus, at different
moments of time in the evolution of the Universe, the ordinary, gravitating, polytropic matter with fixed parameters
$\sigma$ and $n$ creates configurations having different distributions of matter along the radius
(see Fig.~\ref{energ_fig}) and masses (see Fig.~\ref{fig_mass_exp}). On the other hand,
these results may be interpreted as an evolution of the physical characteristics of the same star
whose matter interacts nonminimally with the surrounding chameleon scalar field.

It is important to note that the cosmological time in \eqref{cham_field_evol} is not the same as the Schwarzschild time
of the metric \eqref{metric_sphera}. We are assuming that the cosmological evolution is very slow on the time scale
for the life cycle of a star. Thus, the scalar field is taken to be approximately constant during the life cycle of a given star.
In our analysis the only dynamical time is the time coordinate from \eqref{metric_sphera}. The cosmological time
coordinate from \eqref{cham_field_evol} is given to motivate our use of different (approximately constant) values of the
external scalar field. Of course there could be long-lived stars which have a life cycle long enough so that they would
need to take into account the effect of the change of the scalar field with respect to cosmological time {\it during} the
life cycle of the star. This is a more complex problem, and we will return to it in future work. The general picture that
emerges from the $T^0 _0/ \rho _c c^2$ is that the material of the star is more concentrated at early cosmological times. This
would tend to make these stars hotter and thus have faster life cycles. This fits with our assumption that the scalar field
is approximately constant during the evolution of these early stars. For present times the effect of the scalar field would be
minor. This is born out by the fact that the middle solid line of Fig.~\ref{energ_fig} is very close to the dashed line, which
is the case without any scalar field. The late time stars have a dispersed $T^0 _0/ \rho _c c^2$ which would tend to make the stars
cooler and thus live longer. It is for these late time stars that it might be important to take into account the evolution of the
scalar field during the life of the star.

From Fig.~\ref{energ_fig} one can see what effect the scalar field has on the configuration. For negative values of $\phi_{\text{ext}}$
[which from \eqref{cham_field_evol} correspond to early cosmological times] 
the effect of the presence of the scalar field is that it
tends to concentrate the matter of the star  with respect to the case without a scalar field. 
For small values of $\phi_{\text{ext}}$ [which from \eqref{cham_field_evol} correspond to the present cosmological time] the scalar
field has a minimal effect on the mass-energy distribution of the matter. This can be seen by the very small difference between
the middle solid line in Fig.~\ref{energ_fig} and the dashed line which corresponds to no scalar field at all. Note, even though $\phi_{\text{ext}} =0$
for the middle solid line in Fig.~\ref{energ_fig}, there is a small difference between this situation and having no scalar field, since
even though $\phi_{\text{ext}} =0$ there is still some nonzero distribution of the scalar field inside the star which leads to this small
difference. Finally, for positive values of $\phi_{\text{ext}}$ [which from \eqref{cham_field_evol} correspond to late cosmological times]
the scalar field tends to diffuse the matter of the star. Thus, the general picture that emerges is that the effect of the scalar field at
early times is to concentrate the matter of the star, which would have the effect of making these stars hotter than without the effect of the
scalar field; at the present times the effect of the scalar field would be minimal; at late times the effect would be to disperse the matter of
the star, which would have the effect of making these stars cooler than without the effect of the scalar field.

Figure \ref{fig_mass_exp} shows the dependence of the dimensionless mass $M/M^*$ as a function of $\log_{10}\sigma$
for two values of the polytropic index $n=1.5$ and $n=3.0$, and the values of
the scalar field at the surface of the star, $\phi_1 \simeq \phi_{\text{ext}}=-2, 0, 2$.
Here we have restricted ourselves to just these two values of $n$
since, as in the case of the polytropic sphere without a scalar field considered in \cite{Tooper2},
they provide qualitatively different  behaviors of the mass as a function of $\sigma$.
In the case $n=1.5$, as  opposed to the case $n=3.0$, the mass as a function of $\sigma$ rises to
a maximum and then undergoes damped oscillations about an asymptotic value.

Similar differences in the behavior of the configurations at
$n=1.5$ and $n=3.0$ can be seen from Fig.~\ref{fig_bin_energy_exp}, where the dependence
of the dimensionless binding energy $\text{B.E.}/M^* c^2$ as a function of $\log_{10}\sigma$ is shown.
For $n=1.5$ there is some value $\sigma_{fp}$ of $\sigma$ at which the binding energy has its first peak.
Configurations having $\sigma < \sigma_{fp}$ have positive binding energy, and for
$\sigma > \sigma_{fp}$ the binding energy decreases and eventually becomes negative. In turn, for $n=3.0$
the binding energy is always negative.

As mentioned above, the negativeness of the binding energy implies that the configurations will be
unstable against the dispersal of the fluid of the star to infinity. However, even having a positive binding
energy does not guarantee the stability of a configuration.
For example, with $n=1.5$, and for the configurations considered in \cite{Tooper2},
the binding energy remains positive for $\sigma > \sigma_{fp}$, but the configurations are unstable~--
it is energetically favorable to make a transition to a state with the same
polytropic index $n$, but with a smaller parameter $\sigma$.
Thus, the first mass peak for the case $n=1.5$ from \cite{Tooper2}
corresponds to the point dividing stable and unstable configurations,
and in this case the value $\sigma_{fp}$ corresponds, in Tooper's terminology, to the critical value $\sigma_{cr}$.
For the case of $n=3.0$ there is no peak in either Tooper's model or in the problem being considered here:
all configurations have negative binding energy and are unstable for all $\sigma$ and $\phi_{\text{ext}}$.

The question arises as to whether $\sigma_{fp}$ coincides with $\sigma_{cr}$ in the presence
of the chameleon scalar field for the case $n=1.5$. In \cite{Gleiser:1988rq} this question was studied for a
boson configuration supported by the complex scalar field using the fractional anisotropy of the system.
The fractional anisotropy ($\text{fa}$) is defined by
\begin{equation}
\label{fa}
\text{fa}\equiv (p_r-p_t)/p_r~,
\end{equation}
where the  radial pressure $p_r=-T_1^1$ and the tangential pressure  $p_t=-T_2^2=-T_3^3$ are given by the corresponding energy-momentum tensor.
Using the variational principle suggested by Chandrasekhar \cite{Chandrasekhar:1964zz}, Gleiser \cite{Gleiser:1988rq}
applied the method of trial functions to estimate the dynamical stability of this system.
As a result, contrary to the results for fluid stars, Gleiser found that the instabilities did not occur for central
densities which were equal to or larger than the critical density but for configurations well to the right of the maximum mass.
However,  further studies of such a system showed that the trial functions used in \cite{Gleiser:1988rq} were not the  most general ones,
and that the instability occurred at the  maximum mass, as in the case of ordinary neutron stars (see the erratum \cite{Gleiser:1988rq}).

However, in our system, in addition to the fractional anisotropy, we also have nonminimal coupling between ordinary matter
and the scalar field. Because of this difference, in the next section, we will consider in detail the question of dynamical stability
of the configurations studied here in order to clarify the possible influence of such nonminimal coupling on the stability of the system.
In doing so, the critical value $\sigma_{cr}$ will be calculated by studying the dynamical stability of the configuration
using the approach suggested by Chandrasekhar \cite{Chandrasekhar:1964zz}.
We will not use the method of trial functions used in \cite{Gleiser:1988rq}, but we will instead numerically integrate
Chandrasekhar's pulsation  equations, to trace the behavior of the lowest frequency modes
and thus determine $\sigma_{cr}$.

\section{Dynamical stability}
\label{stab_gen_exp}
\subsection{General equations}

Consider that the equilibrium configurations described in the previous section are perturbed in such a way that spherical
symmetry is maintained. In obtaining the equations for the perturbations, we will neglect all quantities which are
of the second and higher order. The components of the four-velocity are given by \cite{Chandrasekhar:1964zz}
$$
u^0=e^{-\nu_0/2}, \quad u_0=e^{\nu_0/2}, \quad u^1=e^{-\nu_0/2} v, \quad u_1=-e^{\lambda_0-\nu_0/2} v,
$$
with three-velocity
$$
v=\frac{d r}{d x^0} \ll 1 ~.
$$
The index 0 in the metric functions indicates the static, zeroth order solution of the Einstein equations.
The components of the energy-momentum tensor \eqref{emt_cham_star} are then given by
\begin{eqnarray}
\label{emt-00}
&&
T_0^0=f \varepsilon+\frac{1}{2} e^{-\nu}\dot{\varphi}^2+\frac{1}{2} e^{-\lambda}\varphi^{\prime 2}+V(\varphi)
,\\
\label{emt-11}
&&
T_1^1=-f p-\frac{1}{2} e^{-\nu}\dot{\varphi}^2-\frac{1}{2} e^{-\lambda}\varphi^{\prime 2}+V(\varphi),\\
\label{emt-10}
&&
T_0^1=f (\varepsilon+p)u_0 u^1+\partial_0\varphi \partial^1\varphi=
f(\varepsilon_0+p_0)v- e^{-\lambda}\dot{\varphi}\,\varphi^\prime,\\
\label{emt-22}
&&
T_2^2=T_3^3=-f p-\frac{1}{2} e^{-\nu}\dot{\varphi}^2+\frac{1}{2} e^{-\lambda}\varphi^{\prime 2}+V(\varphi).
\end{eqnarray}

Now we consider perturbations of the static solutions of the form
\begin{equation}
\label{perturbations}
y=y_0+y_p ~,
\end{equation}
where the index 0 refers to the static solutions, the index $p$ indicates the perturbation,
and $y$ denotes one of the functions $\lambda, \nu, \varepsilon, p$ or $\varphi$.
Substituting these perturbed expressions into Eqs.~\eqref{Einstein-00_cham_star} and \eqref{Einstein-11_cham_star}, we find
\begin{eqnarray}
\label{Einstein-00pert}
&&
e^{-\lambda_0}\left[r \lambda_p^\prime+\lambda_p\left(1-r \lambda_0^\prime\right)\right]\equiv
\frac{\partial}{\partial r}\left[r e^{-\lambda_0} \lambda_p\right]=
\frac{8\pi G}{c^4} r^2\left[f_0\varepsilon_p+f_p \varepsilon_0
+e^{-\lambda_0}\varphi_0^\prime\left(\varphi_p^\prime-\frac{1}{2}\varphi_0^\prime \lambda_p\right)+V_p\right]
,\\
\label{Einstein-11pert}
&&
e^{-\lambda_0}\left[r \nu_p^\prime-\lambda_p\left(1+r \nu_0^\prime\right)\right]=
\frac{8\pi G}{c^4}r^2\left[f_0 p_p+f_p p_0
+e^{-\lambda_0}\varphi_0^\prime\left(\varphi_p^\prime-\frac{1}{2}\varphi_0^\prime \lambda_p\right)-V_p\right],
\end{eqnarray}
where
$$
f_0=f(\varphi_0), \quad f_p=\varphi_p\left(\frac{d f}{d\varphi}\right)_{\varphi=\varphi_0} , \quad
V_0=V(\varphi_0), \quad V_p=\varphi_p\left(\frac{d V}{d\varphi}\right)_{\varphi=\varphi_0}.
$$
Next, from Eq.~\eqref{Einstein-10_cham_star} we have
\begin{equation}
\label{Einstein-10pert}
-e^{-\lambda_0}\frac{\dot{\lambda}_p}{r}=\frac{8\pi G}{c^4}\left[f_0(\varepsilon_0+p_0)v- e^{-\lambda_0}\dot{\varphi_p}\,\varphi_0^\prime\right].
\end{equation}
Using the components \eqref{emt-00}-\eqref{emt-22} in \eqref{conserv_1_cham_star}, we can rewrite this
equation in terms of the perturbations as
\begin{align}
\label{conserv_osc_pert_gen}
\begin{split}
&-e^{\lambda_0-\nu_0}\left[f_0(\varepsilon_0+p_0)\dot{v}-e^{-\lambda_0}\varphi_0^{\prime}\ddot{\varphi}_p\right]-
\frac{\partial}{\partial r}\left(f_0 p_p+f_p p_0\right)
 \\
&-e^{-\lambda_0}\left\{
\Big(\varphi_0^{\prime\prime}-\lambda_0^{\prime}\varphi_0^{\prime}\Big)
\Big(\varphi_p^\prime-\frac{1}{2}\varphi_0^\prime \lambda_p\Big)+
\varphi_0^\prime\left[\varphi_p^{\prime\prime}-\frac{1}{2}
\left(\varphi_0^{\prime\prime}\lambda_p+\varphi_0^{\prime}\lambda_p^\prime\right)\right]
\right\}+
\frac{\partial V_p}{\partial r}\\
& -\frac{1}{2}\Big[f_0(\varepsilon_p+p_p)+f_p(\varepsilon_0+p_0)\Big]\nu_0^\prime-
\frac{1}{2}f_0(\varepsilon_0+p_0)\nu_p^\prime\\
&-e^{-\lambda_0}\left[\frac{1}{2}\varphi_0^{\prime 2}\nu_p^\prime+\varphi_0^\prime
\left(\varphi_p^\prime-\frac{1}{2}\varphi_0^\prime \lambda_p\right)\nu_0^\prime\right]-
\frac{4}{r}e^{-\lambda_0}\varphi_0^\prime\left(\varphi_p^\prime-\frac{1}{2}\varphi_0^\prime \lambda_p\right)=0.
\end{split}
\end{align}

Now it is useful to introduce a ``Lagrangian displacement'' $\zeta$ with respect to  $x^0$,
$$
v=\frac{\partial \zeta}{\partial x^0}\,.
$$
Equation \eqref{Einstein-10pert} can be directly integrated to give
\begin{equation}
\label{lambda_pert}
\lambda_p=-\frac{8\pi G}{c^4} e^{\lambda_0} r \left[f_0\left(\varepsilon_0+p_0\right)\zeta-e^{-\lambda_0}\varphi_0^\prime \varphi_p\right].
\end{equation}
Inserting this expression into Eq.~\eqref{Einstein-00pert} gives
\begin{equation}
\label{E_pert}
\varepsilon_p=-\frac{1}{f_0}\left\{f_p \,\varepsilon_0+
e^{-\lambda_0}\varphi_0^\prime \left[\varphi_p^\prime-\frac{1}{2}\varphi_0^\prime \lambda_p\right]+V_p+
\frac{1}{r^2}\frac{\partial}{\partial r}\Big[r^2\Big(f_0\left(\varepsilon_0+p_0\right)\zeta-
e^{-\lambda_0}\varphi_0^\prime \varphi_p\Big)\Big]
\right\} ~.
\end{equation}
The expression for $\nu_p^\prime$ follows from Eq.~\eqref{Einstein-11pert},
\begin{equation}
\label{nu_prime_pert}
\nu_p^\prime=\left(1+r \nu_0^\prime\right)\frac{\lambda_p}{r}+
\frac{8\pi G}{c^4} r e^{\lambda_0}
\left[f_0 p_p+f_p p_0+
e^{-\lambda_0}\varphi_0^\prime\left(\varphi_p^\prime-\frac{1}{2}\varphi_0^\prime \lambda_p\right)-V_p\right].
\end{equation}

The perturbed scalar field equation can be found from \eqref{sf_eq_gen}, and is
\begin{align}
\label{phi_pert_gen}
\begin{split}
&\varphi_p^{\prime\prime}- e^{\lambda_0-\nu_0}\ddot{\varphi}_p+
\left[\frac{2}{r}+\frac{1}{2}\left(\nu_0^\prime-\lambda_0^\prime\right)\right]\varphi_p^\prime+
\frac{1}{2}\left(\nu_p^\prime-\lambda_p^\prime\right)\varphi_0^\prime\\
&=e^{\lambda_0}\left\{
\left(\frac{d V}{d\varphi}\right)_p-p_0\left(\frac{d f}{d\varphi}\right)_p-p_p \left(\frac{d f}{d\varphi}\right)_0+
\lambda_p\left[\left(\frac{d V}{d\varphi}\right)_0-p_0 \left(\frac{d f}{d\varphi}\right)_0
\right]
\right\},
\end{split}
\end{align}
where
$$
\left[\frac{d (f,V)}{d\varphi}\right]_0=\left[\frac{d (f,V)}{d\varphi}\right]_{\varphi=\varphi_0}, \qquad
\left[\frac{d (f,V)}{d\varphi}\right]_p=\varphi_p \left[ \frac{d^2 (f,V)}{d\varphi^2}\right]_{\varphi=\varphi_0}.
$$
Thus, we have two linear second-order partial differential equations \eqref{conserv_osc_pert_gen} and \eqref{phi_pert_gen}
for the displacement $\zeta$ and the perturbation of the scalar field $\varphi_p$.
To write out these equations, one needs to use expressions \eqref{lambda_pert}-\eqref{nu_prime_pert}.

\subsection{Case of $f= e^{-b \phi}$}

In this section we apply the equations obtained in the previous subsection to study the stability of the
spherical solution considered in Sec.~\ref{static_sol_exp}. In this case $V=0$, and the equation of state of matter is
taken from expressions \eqref{eqs_cham_star}-\eqref{pressure_fluid_theta}.
From these equations one finds that the perturbed components of the pressure $p_p$ and the
energy density $\varepsilon_p$ are
\begin{equation}
\label{pres_energ_pert}
p_p=K (n+1) \rho_{g c}^{1+1/n}\theta_0^n \theta_p, \quad
\varepsilon_p=n \rho_{g c} c^2 \left[\frac{1}{\theta_0}+\sigma(n+1)\right]\theta_0^n \theta_p.
\end{equation}
In turn, the static components are
\begin{equation}
\label{pres_energ_non_pert}
p_0=K  \rho_{g c}^{1+1/n}\theta_0^{n+1}, \quad
\varepsilon_0= \rho_{g c} c^2 \left(1+\sigma n \theta_0\right)\theta_0^n.
\end{equation}
Next, using expressions \eqref{pres_energ_pert} and \eqref{pres_energ_non_pert} and the dimensionless variables from \eqref{dimless_xi_v}
we write Eqs.~\eqref{lambda_pert}-\eqref{nu_prime_pert} as
\begin{eqnarray}
\label{lambda_pert_app}
&&\lambda_p=-2\sigma (n+1) e^{\lambda_0} \xi \left\{f_0 \left[1+\sigma(n+1)\theta_0\right]\theta_0^n \psi
-e^{-\lambda_0}\phi_0^\prime \phi_p\right\}
,\\
\label{E_pert_app}
&&\theta_p=-\Big\{
n f_0 \Big[\frac{1}{\theta_0}+\sigma(n+1)\Big]\theta_0^n\Big\}^{-1}\nonumber\\
&&\times\Big\{
\Big[
f_p(1+ \sigma n \theta_0)+f_0 \Big[\frac{n}{\theta_0}+\sigma(n+1)^2\Big]\theta_0^{\prime}\psi+
\left[1+\sigma(n+1)\theta_0\right]\Big(f_0 \psi^\prime+\frac{2}{\xi}f_0\psi+\frac{\partial f_0}{\partial \xi}\psi\Big)
\Big]\theta_0^n \nonumber\\
&&-e^{-\lambda_0}\Big[
\Big(\phi_0^{\prime \prime}+\frac{2}{\xi}\phi_0^\prime-\lambda_0^\prime\phi_0^\prime\Big)\phi_p+
\frac{1}{2}\phi_0^{\prime 2}\lambda_p
\Big]
\Big\}
,\\
\label{nu_prime_pert_app}
&&\nu_p^\prime=\left(1+\xi \nu_0^\prime\right)\frac{\lambda_p}{\xi}+
2\sigma (n+1)e^{\lambda_0}\xi
\Big\{
\sigma \Big[
f_0 (n+1)\theta_0^n \theta_p+f_p \,\theta_0^{n+1}
\Big]+e^{-\lambda_0}\phi_0^\prime\Big[
\phi_p^\prime-\frac{1}{2}\phi_0^\prime \lambda_p
\Big]
\Big\},
\end{eqnarray}
where $f_0=e^{-b \phi_0}$, $f_p=-b\, e^{-b \phi_0} \phi_p$, and we have introduced the dimensionless displacement $\psi=A \zeta$.
Equations \eqref{lambda_pert_app}-\eqref{nu_prime_pert_app} will be used to turn the general equations \eqref{conserv_osc_pert_gen}
and \eqref{phi_pert_gen} toward the specific nonminimal coupling under consideration here, namely, $f= e^{-b \phi}$.

To proceed with the stability analysis we assume harmonic perturbations having the following time dependence,
\begin{equation}
\label{harmonic}
y_p(x^0,\xi) = \bar{y}_p(\xi) e^{i\omega x^0}~.
\end{equation}
The function $\bar{y}_p(\xi)$ depends only on the space coordinate $\xi$. For convenience, we hereafter drop the bar.
Using \eqref{harmonic}, the general equation
  \eqref{conserv_osc_pert_gen} takes the form
(we have introduced a new dimensionless frequency $\bar{\omega}=\omega/A$ and we again subsequently drop the bar
for simplicity)
\begin{align}
\label{conserv_osc_pert_exp}
\begin{split}
&\omega^2 e^{\lambda_0-\nu_0}\left[f_0\left[1+\sigma(n+1)\theta_0\right]\theta_0^n \psi
-e^{-\lambda_0}\phi_0^{\prime}\phi_p\right]-
\sigma\frac{d}{d\xi}\left[f_0 (n+1)\theta_0^n \theta_p+
f_p\, \theta_0^{n+1}\right]
 \\
&-e^{-\lambda_0}\left\{
\Big(\phi_0^{\prime\prime}-\lambda_0^{\prime}\phi_0^{\prime}\Big)
\Big(\phi_p^\prime-\frac{1}{2}\phi_0^\prime \lambda_p\Big)+
\phi_0^\prime\left[\phi_p^{\prime\prime}-\frac{1}{2}
\left(\phi_0^{\prime\prime}\lambda_p+\phi_0^{\prime}\lambda_p^\prime\right)\right]
\right\}\\
& -\frac{1}{2}\Big[f_0 \Big(\frac{n}{\theta_0}+\sigma(n+1)^2\Big)\theta_0^n \theta_p
+f_p\Big(1+\sigma(n+1)\theta_0\Big)\theta_0^n
\Big]\nu_0^\prime-
\frac{1}{2}f_0\Big(1+\sigma(n+1)\theta_0\Big)\theta_0^n \nu_p^\prime\\
&-e^{-\lambda_0}\left[\frac{1}{2}\phi_0^{\prime 2}\nu_p^\prime+\phi_0^\prime
\left(\phi_p^\prime-\frac{1}{2}\phi_0^\prime \lambda_p\right)\nu_0^\prime\right]-
\frac{4}{\xi}\,e^{-\lambda_0}\phi_0^\prime\left(\phi_p^\prime-\frac{1}{2}\phi_0^\prime \lambda_p\right)=0.
\end{split}
\end{align}
Next, the perturbation equation for the scalar field \eqref{phi_pert_gen} is given by
\begin{align}
\label{phi_pert_exp}
\begin{split}
&\phi_p^{\prime\prime}+
\left[\frac{2}{\xi}+\frac{1}{2}\left(\nu_0^\prime-\lambda_0^\prime\right)\right]\phi_p^\prime+
\frac{1}{2}\left(\nu_p^\prime-\lambda_p^\prime\right)\phi_0^\prime
+ \omega^2 e^{\lambda_0-\nu_0}\phi_p\\
&=-\sigma e^{\lambda_0}\left\{
b^2 f_0 \theta_0^{n+1}\phi_p-b f_0 \left[(n+1)\theta_p+\theta_0\lambda_p\right]\theta_0^n
\right\}.
\end{split}
\end{align}
Thus, we have two linear second-order ordinary differential equations \eqref{conserv_osc_pert_exp} and \eqref{phi_pert_exp}
for the dimensionless displacement $\psi$
and the perturbations of the scalar field $\phi_p$. Equation \eqref{conserv_osc_pert_exp} is an analog of
Chandrasekhar's ``pulsation equation'' \cite{Chandrasekhar:1964zz}. Solutions of this equation must satisfy the boundary conditions
$\psi(0)=0$  and $p_p=0$ at $\xi=\xi_1$. On the other hand, since the total expression for the perturbed radial pressure
 $\delta p_r=-\delta T_1^1$ [with $T_1^1$ taken from \eqref{emt-11}] contains not only the fluid component $p_p$
but also perturbations of the scalar field, it is necessary to require that the total perturbation
$\delta p_r$ be equal to zero as well. Formally, the scalar field energy goes to zero at infinity, i.e. as $\xi \to \infty$.
Thus, the boundary conditions for the system of coupled equations \eqref{conserv_osc_pert_exp} and \eqref{phi_pert_exp} are
\begin{equation}
\label{bound_cham_exp}
\psi(0)=\phi_p(0)=0, \quad \delta p_r \to 0 \quad \text{as} \quad \xi \to \infty.
\end{equation}
The system \eqref{conserv_osc_pert_exp} and \eqref{phi_pert_exp}, together with the boundary conditions \eqref{bound_cham_exp},
defines a characteristic value problem for $\omega^2$. The question of stability is thus reduced to a study of the possible
values of  $\omega^2$. If any of the values of $\omega^2$ are found to be negative, then the perturbations will grow and the
configuration is unstable against radial oscillations. Since the eigenvalues form a discrete sequence, our purpose is to
find the lowest eigenvalue $\omega_0^2$, and if $\omega_0^2 >0$ the configuration is stable.

Of course, the value of $\omega_0^2$ depends on the values of the parameters  $n, \sigma, b$, and the value
of the background scalar field $\phi_{\text{ext}}$. Here we
 find the values of $\sigma$ for which the
equilibrium configurations are stable at fixed $n, b$, and $\phi_{\text{ext}}$. This will allow us to determine
the boundary between stable and unstable configurations. This boundary is defined by the point where $\omega_0^2$, as
function of $\sigma$, crosses through zero. We denote the corresponding critical values of  $\sigma$ as $\sigma_{cr}$.

To find $\sigma_{cr}$, we numerically solved Eqs.~\eqref{conserv_osc_pert_exp} and \eqref{phi_pert_exp}
to determine the characteristic values for the frequency $\omega^2$.
In the absence of the scalar field, both the energy approach to stability and the dynamical stability analysis give the
same values of $\sigma_{cr}$ \cite{Tooper2}. Further, without a scalar field, $\sigma_{cr}$ is given by
the first peak
in both the mass as a function of $\sigma$ and the binding energy as a function of $\sigma$ distributions. However,
in the presence of a scalar field the point dividing stable and unstable configurations may shift.
Therefore, in  finding $\sigma_{cr}$,  we proceed as follows: we choose the initial value of
$\sigma$ corresponding  to the first peak in the binding energy (or in the mass) functions.
For this value, we find  the lowest $\omega_0^2$.
We always get positive values of  $\omega_0^2$ at the first peak, which means that
these configurations are stable. Next, shifting to the right from the first peak, we look for the value of $\sigma$
at which $\omega_0^2=0$, which gives $\sigma_{cr}$.

\begin{figure}[t]
\centering
  \includegraphics[height=7cm]{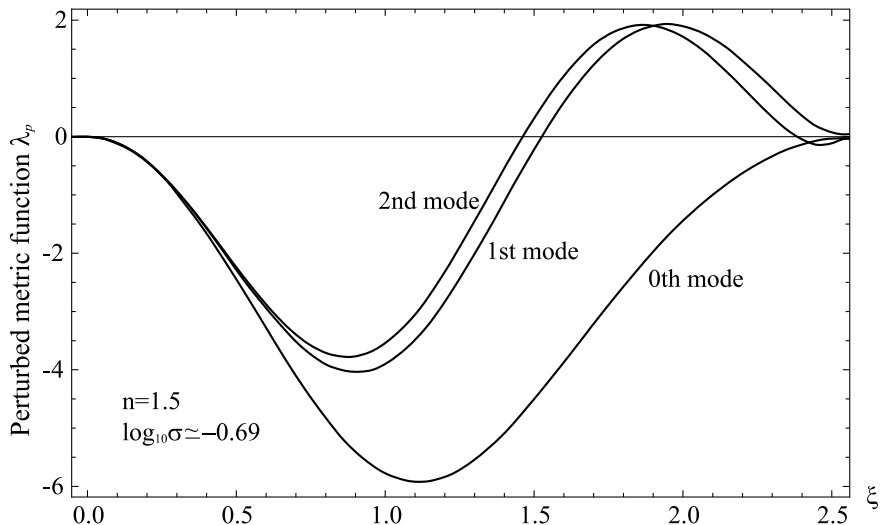}
\caption{The perturbation in the metric function $\lambda_p$ is shown as a function of the radial coordinate $\xi$ for the
first three modes for the critical value $\log_{10}\sigma_{cr}\simeq -0.69$. The curves are plotted for $b=1$ and $\phi_{\text{ext}}= 0$.}
\label{eigen_lambda}
\end{figure}

We applied the above procedure for finding  $\sigma_{cr}$ for the configurations with the fixed
values of $n, b$, and $\phi_{\text{ext}}$ used in Sec.~\ref{num_res}. The values of
$\sigma_{cr}$ are shown in Figs.~\ref{fig_mass_exp} and \ref{fig_bin_energy_exp} by the bold dots. The dots
mark the dividing point between stable configurations (to the left of the dots) and unstable ones (to the right of the dots).
As expected, for the case $n=3.0$, when the binding energy is always negative,
there are only negative values of  $\omega_0^2$; i.e. the configurations are unstable at any $\sigma$. Thus, in 
Figs.~\ref{fig_mass_exp} and \ref{fig_bin_energy_exp}, only the results for the case of $n=1.5$ are given. In this case the numerical calculations give
$\log_{10}\sigma_{cr} \simeq -0.69$ for all values  $\phi_{\text{ext}}=-2, 0, 2$.
In turn, the fractional anisotropy \eqref{fa} is given by
$$
\text{fa}=\frac{\left[1-2\sigma(n+1)v_0/\xi\right]\phi_0^{\prime 2}}
{\sigma f_0 \theta_0^{n+1}+1/2 \left[1-2\sigma(n+1)v_0/\xi\right]\phi_0^{\prime 2}}\,.
$$
It can be found from this expression that at the center of the configurations
$\text{fa}(0)=0$, and at the boundary,  $\xi=\xi_1$, $\text{fa}(\xi_1)$  is always equal to 2.

The eigenfunctions corresponding  to the lowest $\omega_0^2$ have  no  nodes,
and the  eigenfunctions  corresponding  to  the $n$th  mode  have  $n$  nodes. As an example, Fig.~\ref{eigen_lambda} shows the plots
of the eigenfunction $\lambda_p$ from  \eqref{lambda_pert_app} for the above value  $\log_{10}\sigma_{cr} \simeq -0.69$.
Using Eqs.~\eqref{conserv_osc_pert_exp} and \eqref{phi_pert_exp} and the boundary conditions \eqref{bound_cham_exp},
we plot in this figure the first three modes. For the zero mode  $\omega_0^2=0$, for the first mode $\omega_1^2 \simeq 0.18$,
and for the second mode $\omega_2^2 \simeq 0.22$.

\section{Conclusion}
\label{concl_exp}

Based on the assumption that   time evolving chameleon  scalar fields may exist in the Universe,
this paper studies the possible influence that such fields may have
on the inner structure and stability of compact gravitating configurations
consisting of ordinary matter and embedded in an external, homogeneous chameleon  scalar field.
The ordinary matter is taken to have a polytropic equation of state which is
parametrically given by expressions \eqref{eqs_cham_star}-\eqref{pressure_fluid_theta}.
We choose a specific form for the nonminimal coupling between the ordinary matter and the scalar field
 given  by \eqref{fun_f_exp}. This form was used
in \cite{Farajollahi:2010pk} within the framework of chameleon cosmology.
Here we study the consequences  of the presence of a chameleon scalar field
on the inner structure of compact astrophysical objects (i.e. ``stars")
rather than the cosmological consequences of chameleon fields.
In Ref.~\cite{Dzhunushaliev:2011ma} this type of gravitating configuration
of a polytropic fluid nonminimally coupled to a scalar field was called a ``chameleon star.''

One of the main conclusions of the present work is that the characteristics of the chameleon stars  --
their masses, sizes, matter distribution -- depend strongly on  the parameters of the surrounding scalar field $\phi$.
Assuming that  this background scalar field changes with cosmological times, we predict that the
characteristics of the stars embedded in the scalar field should also change.
For example, we considered three different values of the background scalar field,
$\phi_{\text{ext}}=-2, 0, 2$, which could represent the time evolved value of the field
at different times according to the law \eqref{cham_field_evol}. These three values
could correspond to three different periods of time in the evolution of the Universe -- from the relatively early Universe to
the future Universe, for example, already  near the big rip singularity.
The results of the numerical calculations presented in 
Figs.~\ref{energ_fig}-\ref{fig_bin_energy_exp} (which cover the values of the polytropic index, $n=1.5$ and $n=3.0$)
indicate that the characteristics of the configurations have a strong dependence on the value of the scalar field $\phi_{\text{ext}}$.
In particular, from Fig.~\ref{energ_fig} one finds that, at the center of the configurations considered,
there can be either a larger or smaller concentration of the matter density compared to the case of a star without a scalar field.
This would have an impact  on the rate of fusion reactions inside the stars, and thus on their lifetimes so that
the life span of the star could be altered by the properties of the background chameleon scalar field surrounding the star.
Some other speculations about possible physical applications of the chameleon star model can be found
in Ref.~\cite{Dzhunushaliev:2011ma}.

Another important consequence of the presence of the chameleon scalar field
is its influence on the stability of the configurations studied here. From previous investigations of
the stability of polytropic gravitating fluids it is known that for equilibrium configurations
instabilities occur once one moves to the right of the first peak in the binding energy (or the first peak in the mass) \cite{Tooper2}.
A similar situation also occurs for purely boson stars (see the erratum from Ref.~\cite{Gleiser:1988rq}).
However, the dynamical stability analysis performed in Sec.~\ref{stab_gen_exp}
shows that the configurations considered here have their stable points shifted to the right of the first peak in the
binding energy. These shifts are shown by the bold dots in Figs.~\ref{fig_mass_exp} and \ref{fig_bin_energy_exp}.
Thus, in contrast to the usual polytropic and boson stars, 
the  expansion  of  the
region of stability is in the direction where $\sigma$ increases.
It is obvious that this new effect
is caused by the presence in the system of the nonminimal coupling between the ordinary matter and the scalar field.

The study of mixed configurations with scalar fields and ordinary matter sources is of interest
since various types of scalar fields are believed to exist in the Universe.
Most often, such scalar fields play a cosmological role~-- as a mechanism to drive inflation (the inflaton
field), as dark matter, or as dark energy. Here we consider the possibility that one type of scalar field
(a chameleon scalar field) might also have some influence on astrophysical objects such as stars.

\section*{Acknowledgements}
V.F. is grateful to the Research Group Linkage
Programme of the Alexander von Humboldt Foundation for the
support of this research. He also would like to thank the
Carl von Ossietzky University of Oldenburg for hospitality
while this work was carried out.

\end{document}